\documentclass[10pt,twocolumn]{article}
\usepackage[cp1251]{inputenc}
\usepackage[english]{babel}
\usepackage{graphicx}
\usepackage{amssymb}
\usepackage{amsmath}
\usepackage{amsthm}
\usepackage{amsfonts}
\usepackage{amssymb}

\title{{\bf \Large  Testing Fractional Action Cosmology}\\
{\normalsize ~~{\bf V.\,K. Shchigolev}\thanks{E-mail:
vkshch@yahoo.com}}\\ \vspace{5mm}
{\small {\it Department of Theoretical Physics, Ulyanovsk State University, \\ \vspace{-2mm}42 L. Tolstoy Str.,
Ulyanovsk 432000, Russia}}\\
\vspace{2mm}
\small \begin{quote}{\bf Abstract} ~~The present work deals with a combined test of the so-called Fractional Action Cosmology (FAC) on the example of a specific model obtained by the author earlier. In this model, the effective cosmological term is proportional to the Hubble parameter squared through the so-called kinematic induction. The reason of studying this cosmological model could be explained by its ability to describe two periods of accelerated expansion, that is in agreement with the recent observations and the cosmological inflation paradigm.  First of all, we put our  model through the theoretical tests that gives a general conception of the influence of the model parameters on its behavior. Then, we obtain some restrictions on the principal parameters of the model, including the fractional index, by means of the observational data. Finally, the cosmography parameters and the observational data compared to the theoretical predictions are presented both analytically and graphically.
 \\
\vspace{2,5mm}
{\bf PACS numbers}: 98.80.-k; 98.80.Jk; 04.20.Jb\\
{\bf Key words}:  fractional Einstein-Hilbert action, cosmological models, accelerated expansion,  kinematics tests, observational constraint\\
\end{quote}}
\date{}
\usepackage[left=30mm, right=20mm, top=25mm, bottom=25mm]{geometry}

\begin{document}
\maketitle
\vspace{-20mm}
\section{Introduction}
\quad
According to the recent cosmological observations, including the type Ia supernova, the cosmic microwave background radiation, and the large-scale structure, one could be assured that our Universe is undergoing the accelerated expansion \cite{Riess1}-\cite{Allen}.
The rapid growth of the number of publications for the last decade and a half, devoted to the mysteries of the late cosmological acceleration, led to a radical revision of cosmological models. Among these modifications, two main trends should be noted. First of all, a great amount of models with often exotic forms of matter called commonly Dark Energy (DE) which equation of state (EoS) satisfies $w= p/\rho< -1/3$ has been developed. Various kinds of DE models have been proposed such as a cosmological constant \cite{Peebles}, quintessence \cite{Caldwell1,Sami},  phantom \cite{Caldwell2}-\cite{Cline}, tachyon \cite{Sen1}, \cite{Gibbons}, k-essence \cite{Mukhanov}-\cite{Scherrer}, Chaplygin gas \cite{Kamenshchik}, quintom \cite{Feng}, holographic dark energy \cite{Horava} and Yang--Mills fields \cite{0Shchigolev}, \cite{00Shchigolev}, etc.

The alternative approach to the problem of accelerated expansion
refers to the consideration of a numerous modifications  of the gravity theory.  Among various modifications, the multidimensional theory, braneworld models, teleparallel theory and many others could be mentioned \cite{Sahni1}-\cite{Setare}.  Nevertheless, the puzzle of the present cosmological acceleration still remains one of the main  problems in modern cosmology.

At the same time, the observational data and our knowledge presume the paradigm of cosmological inflation regarding the very early universe. Therefore, any reliable cosmological model should consist of at least two periods of accelerated expansion.  In this respect, it is interesting to consider any analytic models, which are capable of, more or less realistically, describe the whole evolution from (almost) the beginning to the present day \cite{Nojiri1} -\cite{4Shchigolev}.

It should be noted that a phenomenological approach in theoretical cosmology is considered as quite applicable one in the recent years  due to the great problem in description of the behavior of our Universe which is followed from the observed phenomenon of the late-time cosmological acceleration. Indeed, one could recall some classes of modified gravities containing $f(R)$, $f(G)$, $f(R, G)$ and $F(R, T)$ which are considered to be gravitational alternatives
for DE \cite{Starobinsky1}- \cite{Myrzakulov}.

Among phenomenological models, a model based on the ideas of fractional calculus seems to be quite attractive in its ability to describe two periods of acceleration \cite{3Shchigolev}.
Based on the concept of fractional calculus of variations (or the fractional action-like variational approach), the fractional action cosmology  with fractional weight function has been proposed recently in \cite{1Nabulsi}-\cite{3Nabulsi}. These theories are derived from the action integral, which is built on  the  Lebesgue-Stieltjes measure $d \varrho(x)$ generalizing the standard 4-dimensional measure $d^4x$. In FAC,  the action integral $S_L [q]$ for the Lagrangian density $L(t', q_i(t'), \dot q_i(t'))$ is written as a fractional Riemann-Stieltjes integral: $S^{\alpha}_L [q_i]=\Gamma^{-1}
(\alpha)\int\limits_{t_0}^{t} L(t',q_i (t'),\dot
q_i(t'))(t-t')^{\alpha-1} d t'$ with the integrating function ${\displaystyle g_t(t') = \Gamma^{-1}
(\alpha) [t^{\alpha} - (t-t')^{\alpha}]}$. This approach realizes  the space
scaling concepts of Mandelbrot to define the scaling in fractional time as
$d^{\alpha}t = \pi^{\alpha/2} \Gamma^{-1}(\alpha/2) |t|^{\alpha-1} dt$, where $0<\alpha <1$ \cite{Mandelbrot}.

Several interesting features of FAC have inspired a series of works studying such a kind of theories  for the last few years (see, e.g.,  \cite{1Shchigolev}  - \cite{4Nabulsi} and references therein). One of the features of FAC is that the continuity equation perturbs  the energy-momentum conservation law. One could mention that a perturbed continuity equation is not some specific property of the fractional (and fractal) cosmology but is almost common feature of many modifications of the gravity theory. The main idea of our work \cite{3Shchigolev} is keeping the usual form for the continuity equation  in the framework of FAC. We  achieved this aim by means of retaining the concept of fractional order for the action functional only with respect to the gravity sector. Moreover, we proposed to arrange the set of main equations in such a way  that the effective  $\Lambda$ - term could be treated as a kinematically induced (by the Hubble parameter) cosmological term.   We showed in a specific example that the model based on this proposal  could lead to some rather realistic regimes of  expansion of the universe.

On this background, it is particularly important to put such a model through the kinematic  and observational tests in order to determine whether it can describe the real behavior of the Universe. At the same time, one could try to obtain the numerical values of the fractional index $\alpha$ and other parameters of the model that fit well the observed characteristics of the evolving Universe.

In this work, we study a model in which the effective cosmological term $\Lambda_{eff}\propto H^2$, providing two accelerated periods in the evolution of the Universe. In this regard, we
consider a spatially flat Friedmann-Robertson-Walker (FRW) background equipped with such a time-varying $\Lambda_{eff}(t)$.  To testify this model, we consider some theoretical and observational tests.
First, we get a first hint on the possible values of the fractional parameter $\alpha$ reproducing the cosmographic parameters for our model.
Second, with the aim to constrain our model by the observational data,  we estimate $\alpha$  and other parameters with the help of $H(z)$ data. This is a minimal approach but it is useful in order to probe the selfconsistency of the model. Then, the possible observational restrictions of our cosmological model are determined through the SN Ia Union2 database.

\section{Fractional Einstein-Hilbert Action Cosmology}
\quad
In our previous work \cite{3Shchigolev}, we considered a FAC model derived from a variational principle for the modified fractional Einstein-Hilbert action with a varying cosmological term $\Lambda$, that is  $\displaystyle S^{\alpha}_{EH}=M_{P}^2\int \sqrt{-g}\,g_{t'}(t)\,d^{\,4}x (R-2\Lambda)/2$ where $M_P^{-2}=8\pi G$ is reduced Planck mass.
Assuming the matter content of the universe is minimally coupled to gravity, the total action of the system is $S_{total}^{\alpha}=S_{EH}^{\alpha}+S_{m}$, where the effective action for matter can be represented by the usual expression, which follows from the matter action  with the standard measure $S_{m}=\int {\cal L}\sqrt{-g} d^4x$.

Using the fractional variational procedure in a spatially flat FRW metric, $\displaystyle ds^2 = dt^2 - a^2(t)\delta_ {ik} dx^i dx^k$, where $a(t)$ is a scale factor,  we obtain the following dynamical equations for our model:
\begin{equation}
3 H^2+3\frac{(1-\alpha)}{t}H=t^{1-\alpha} \rho+\Lambda ,\label{1}
\end{equation}
\begin{equation}
2\dot H + 3 H^2+\frac{2(1-\alpha)}{t}H +
\frac{(1-\alpha)(2-\alpha)}{t^2}\!=\!-t^{1-\alpha} p+\Lambda ,\label{2}
\end{equation}
where $H(t)=\dot a/a$ is the Hubble parameter, and  we set $8\pi G\, \Gamma
(\alpha)=1$ for simplicity. The continuity equation is written in its usual form,
\begin{equation}
\dot \rho+3H(\rho+p)=0,\label{3}
\end{equation}
expressing the standard energy conservation law for a perfect fluid, just the same
as in the cosmological theory of General Relativity.
From Eqs. (\ref{1}) and (\ref{2}), one can obtain the EoS of matter as follows:
\begin{equation}\label{4}
w_m = \frac{p}{\rho}= -1-\frac{2}{3}\,\frac{\displaystyle \frac{\dot H}{H^2}-\frac{1-\alpha}{2(tH)}+\frac{(1-\alpha)(2-\alpha)}{2(tH)^2}}{\displaystyle 1+\frac{1-\alpha}{(tH)}-\frac{\Lambda}{3 H^2}}.
\end{equation}

It can be shown that Eqs. (\ref{1}) and (\ref{2}) imply the continuity equation (\ref{3}) in the case $\alpha \ne 1$, only if the following equation is valid:
\begin{equation}
\frac{d }{d t}\Big(t^{\alpha - 1}\Lambda\Big)=\frac{3(1-\alpha)}{t^{ 2-\alpha}}\,\Big[\dot H - 2\frac{(2-\alpha)}{t}H\Big].\nonumber
\end{equation}
This equation can be solved in quadratures as
\begin{eqnarray}
\Lambda(t)=\Lambda_0 t^{1-\alpha}+~~~~~~~~~~~~~~~~~~~~~ \nonumber \\ +3(1-\alpha)\left[\frac{H(t)}{t}-(2-\alpha)t^{1-\alpha}\int t^{\alpha-3} H(t)d t\right], \label{5}
\end{eqnarray}
where $\Lambda_0$  is a constant of integration. Substituting Eq. (\ref{5}) into the basic equations of our model (\ref{1}), (\ref{2}), we obtain the following set of equations:
\begin{equation}
3H^2=t^{1-\alpha}\rho_{eff},\label{6}
\end{equation}
\begin{equation}
2\dot H+3H^2-\frac{1-\alpha}{t}H+\frac{(1-\alpha)(2-\alpha)}{t^2}=-t^{1-\alpha}p_{eff},\label{7}
\end{equation}
where the effective energy density and pressure are represented by
\begin{equation}
\rho_{eff}=\rho+\Lambda_{eff},\,\,\,\,\,\,p_{eff}=p-\Lambda_{eff},\nonumber
\end{equation}
\begin{equation}
\Lambda_{eff}=\Lambda_0-3(1-\alpha)(2-\alpha)\int t^{ \alpha-3}H(t)d\,t.\label{8}
\end{equation}
The last equation supposes  that the effective cosmological term is the sum of the cosmological constant $\Lambda_0$ and induced  cosmological term $\Lambda_{ind}$.

\begin{figure}[thbp]
\centering
\includegraphics[width=0.45\textwidth,height=0.45\textwidth]{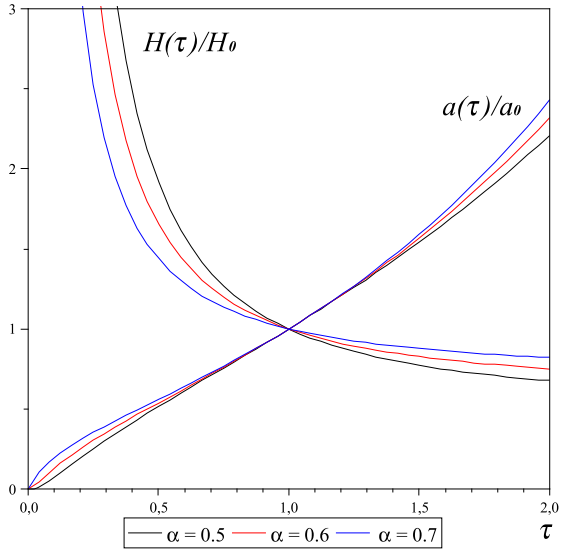}
\caption{Evolution of the $\Lambda_{eff} = \beta H^2$ model for several values of the fractional index  $\alpha$.}
\label{Figure_1}
\end{figure}

Thus, the set of dynamical equations (\ref{6}), (\ref{7}) consists of two independent equations, and fully determines the dynamics of our model. However, to determine three parameters,  one more condition should be set. For example,  an effective EoS can play the role of such  additional equation. Indeed, if we consider the effective barotropic fluid, then the effective EoS  follows from Eqs. (\ref{6}) and (\ref{7}) in the form:
\begin{equation}\label{9}
w_{eff} = \frac{p_{eff}}{\rho_{eff}} = -1 -\frac{2}{3}  \frac{\dot H}{H^2}+\frac{1-\alpha}{3(tH)}\left[1-\frac{2-\alpha}{(tH)}\,\right].
\end{equation}
Assuming the matter obeys also a barotropic EoS $p= w_m \rho$, we obtain the following equations relating the barotropic indexes of the effective fluid and matter:
\begin{eqnarray}w_m=\frac{p}{\rho}=-1 +\frac{ 1+w_{eff}}{\displaystyle 1-\frac{\Lambda_{eff}}{3H^2}t^{ 1-\alpha}},\nonumber\\
w_{eff}=w_m -(1+w_m)\frac{\Lambda_{eff}}{3H^2}t^{ 1-\alpha}.\label{10}
\end{eqnarray}

However, this approach requires a specification of the matter content of the universe or some hypotheses concerning the behavior of the effective EoS. Consideration of hypothetical forms of the dark matter and DE causes the most problems in modern cosmology. Therefore, we are going to consider an alternative approach to our model.
In this approach, to make the system of equations (\ref{6}), (\ref{7})
closed i.e to make the number of unknowns and number
of equations equal, we suppose that $\Lambda_{eff}$ is given as
a function of other parameters of the model.
To determine the evolution of our model, we can supplement the system of equations (\ref{6}), (\ref{7}) by the following effective cosmological term
\begin{equation}\label{11}
\Lambda_{eff} = \beta H^2,
\end{equation}
which is often discussed in the literature (see e.g.  \cite{Overduin}, \cite{Sahni2}).
For the convenience of further calculations and simplification of all equations, we assume a constant of proportionality $\beta$ in (\ref{11}) is given by
\begin{equation}\label{12}
\beta =\frac{3}{2} m H_0^{1-\alpha},
\end{equation}
where $m$ is a new constant, and $\alpha \in (0,1)$.
Solving Eq. (\ref{8}) together with (\ref{11}), we obtain
\begin{equation}\label{13}
H(\tau)=H_0\left[\frac{1-\alpha}{m \tau^{2-\alpha}}+\frac{m-(1-\alpha)}{m}\right],
\end{equation}
where we have introduced the dimensionless cosmic time $\tau = H_0 t$ so that $H(\tau=1)=H_0 $. Therefore, in the far future, the Hubble parameter reaches the following value
$$
H_{\infty}=H(t \to \infty)=\frac{ (m-1+\alpha)}{m}H_0,
$$
and then $\Lambda_0=\beta H_{\infty}^2$. From equation (\ref{13}), the scale factor of this model is equal to
\begin{equation}\label{14}
a(\tau) \!=\! a_0 \exp \left[\Big(1\!-\!\frac{1-\alpha}{m}\Big)\tau\!-\!\frac{1}{m \tau^{ 1-\alpha}}\!+\!\frac{2\!-\!\alpha-m}{m}\right],
\end{equation}
where $a_0 = a(\tau=1)$. The red shift $z$ is defined as $1+z=a_0/a(\tau)$. Therefore, one can obtain from equation (\ref{14}) that
\begin{equation}\label{15}
z\!=\! \exp \left[\frac{1}{m \tau^{1-\alpha}}-\Big(1-\frac{1-\alpha}{m}\Big)\tau-\frac{2-\alpha-m}{m}\right]-1.
\end{equation}
The evolution of this model according to Eqs.  (\ref{13}) and (\ref{14}) for the specific value of parameter $m=1$ is shown in Fig. 1.

\begin{figure}[thbp]
\centering
\includegraphics[width=0.45\textwidth,height=0.45\textwidth]{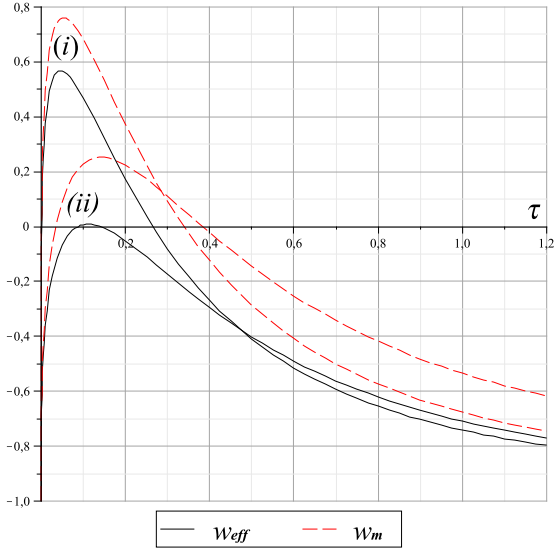}
\caption{Graphs of the effective EoS and the EoS of matter versus time for (i) $\alpha=0.800,m=0.400$; and (ii) $\alpha=0.684,m=0.749$.}
\label{Figure_1}
\end{figure}

Making use of Eqs. (\ref{9}) and (\ref{13}),  it is easy to obtain the effective EoS
\begin{eqnarray}
w_{eff}(\tau)\! =\! -1\!+\!\frac{m (1\!-\!\alpha)\,\tau^{ 1-\alpha}}{3\Big[m(1\!-\!\alpha)+(m\!-\!1+\alpha)\tau^{ 2-\alpha}  \Big]} \nonumber \\
\times \left[\!1\!+\!\frac{m(2\!-\!\alpha)(2-\tau^{ 1-\alpha})}{m(1\!-\!\alpha)+(m\!-\!1\!+\!\alpha)\tau^{ 2-\alpha}}\right],\label{16}
\end{eqnarray}
and the EoS of matter
\begin{equation}\label{17}
w_m(\tau)=-1+\frac{2[1+w_{eff}(\tau)]}{2-m\tau^{\displaystyle 1-\alpha}},
\end{equation}
where $w_{eff}$ is represented by Eq. (\ref{16}). The evolution of the effective EoS and the EoS of matter in  $\Lambda_{eff} = \beta H^2$ model according to Eqs.  (\ref{16}) and (\ref{17}) are depicted in Fig. 2.
This Figure demonstrates that the effective EoS $w_{eff}$
starts its growing  at  $w_{eff}(0)=q(0)=-1$ at $\tau=0$
and asymptotically approaches to the same value:  $w_{eff}(t \to \infty)=q(t \to \infty)=-1$.
However, for a certain period, the effective EoS $0<w_{eff}<1$.

\begin{figure}[h]
\centering
\includegraphics[width=0.45\textwidth,height=0.45\textwidth]{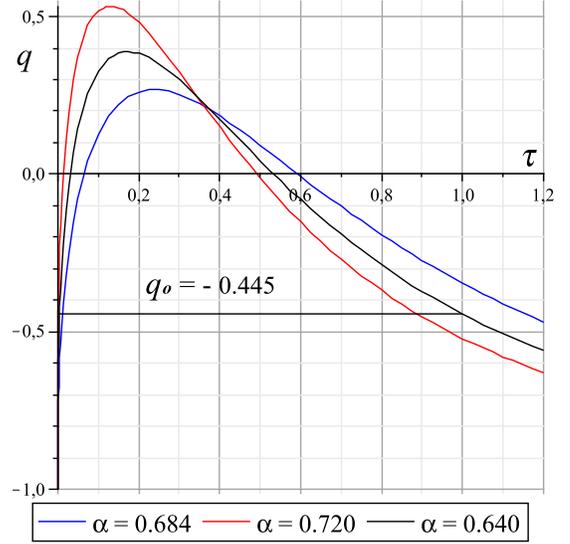}
\caption{Shows the deceleration parameter $q$ versus cosmic time $\tau$. Here, $m=0.749$.}
\label{Figure_3}
\end{figure}

An important observational quantity is the deceleration
parameter defined by $q = -a^2\, \ddot a/\dot a^2 = -1-\dot H/H^2$
where the sign of $q$ indicates whether the model accelerates
or not. For a decelerating model we have $q > 0$,
whereas for an accelerating model of the universe
$q < 0$. The deceleration parameter in FAC is defined just as in the standard cosmology, as it is a cosmography parameter of the model.
Due to equation (\ref{13}), it can be expressed by
\begin{equation}\label{18}
q(\tau)=-1+\frac{m(1-\alpha)(2-\alpha)\tau^{1-\alpha}}{[1-\alpha+(m-1+\alpha)\tau^{2-\alpha}]^2}.
\end{equation}
Fig. 3 shows the evolution of $q$ as a function of cosmic time $\tau$. It is interesting to notice that for $\alpha \approx 0.684$ and $m=0.749$, the present deceleration parameter is $q_0=q(\tau=1) \approx -0.445$. Let us note that these values of parameters will be obtained from the observational data in what follows.

\section{Theoretical Tests and Observational Constraints}
\quad
Obviously, the tuning of our model is possible by means of several parameters, such as $\alpha, m$ and $H_0$. In this section, we explore several diagnostics for our model based on theoretical background and on the observational data.

\subsection{Statefinder Diagnostics}
\quad
In order to distinguish between  various DE models, V. Sahni et al. \cite{Starobinsky} proposed a cosmological diagnostic pair $\{r, s\}$ called statefinder.
The statefinder test is a geometrical one based on the expansion of the scale factor a(t) near the present time $t_0$:
$$
a(t)/a(t_0) = 1 + H_0 (t - t_0) -\frac{1}{2} q_0 H^2_0 (t - t_0)^2
$$
$$
+\frac{1}{6}r_0 H^3_0 (t - t_0)^3 +\frac{1}{24}s_0 H^4_0 (t - t_0)^4 +\mathcal{O}([t-t_0]^5,
$$
where $H_0, q_0, r_0$ are the present values of the Hubble parameters, deceleration parameter and the statefinder indexes $r = \dddot a /a H^3$ and $s = \ddddot a /a H^4$ respectively. The statefinder parameter $s$ is the combination of $r$ and $q$: $s = (r - 1)/3(q - 1/2)$.
The important feature of statefinder is that the spatially flat $\Lambda$CDM has a fixed point
$\{r, s\} = \{1, 0\}$. Departure of a DE model from this fixed point is a good way of establishing the ‘distance’ of this model from flat $\Lambda$CDM. In terms of the Hubble parameter and its derivatives with respect to cosmic time, the statefinder parameters of a flat FRW model are given by
\begin{equation}\label{19}
r=1+3\frac{\dot H}{H^2}+\frac{\ddot H}{H^3},\,\,\,s= - \Big(\frac{2}{3 H}\Big)\,\frac{3 H \dot H+\ddot H}{3 H^2+2 \dot H}\,.
\end{equation}
With the help of (\ref{13}) and (\ref{19}), one can get that
\begin{equation}
r(\tau)=1-m(1-\alpha)(2-\alpha)\tau^{1-\alpha}\nonumber
\end{equation}
\begin{equation}\label{20}
\times\frac{[3(1\!-\!\alpha)\!-\!m(3\!-\!\alpha)\tau^{1-\alpha}+
3(m\!-1\!+\!\alpha)\tau^{2-\alpha}]}{[1-\alpha+(m-1+\alpha)\tau^{2-\alpha}]^3},
\end{equation}
and
\begin{equation}
s(\tau)=\frac{2m(1-\alpha)(2-\alpha)\tau^{1-\alpha}}{3[1-\alpha+(m-1+\alpha)\tau^{2-\alpha}]} \nonumber
\end{equation}
\begin{equation}\label{21}
\times\frac{3(1-\alpha)-m(3-\alpha)\tau^{1-\alpha}+
3(m-1+\alpha)\tau^{2-\alpha}}{3[\!1\!-\!\alpha\!+\!(m\!-\!1\!+\!\alpha)
\tau^{2-\alpha}]^2\!\!-\!2m(1\!-\!\alpha)(2\!-\!\alpha)\tau^{1-\alpha}}.
\end{equation}
It can be easily seen from equations (\ref{20}) and (\ref{21}) that our model is able to appear in the $\Lambda$CDM point $\{r, s\} = \{1, 0\}$ at the beginning moment $\tau=0$, and at $\tau \to \infty$. The curves on $\{r, s\}$ plan for some different values of parameter $\alpha$ and $m=1$ are shown in Fig. 4.
\begin{figure}[thbp]
\centering
\includegraphics[width=0.45\textwidth,height=0.45\textwidth]{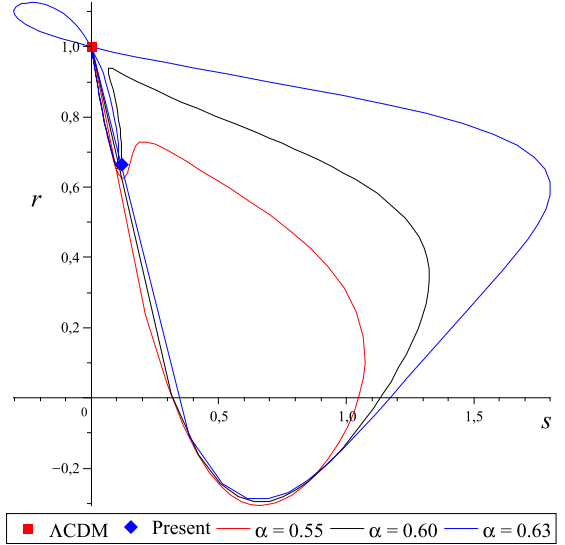}
\caption{The lines on $\{r, s\}$ plan for different values of parameter $\alpha$. The $\{1, 0\}$ point is related to the location of standard $\Lambda\mbox{CDM}$ model.}
\label{Figure_4}
\end{figure}

\subsection{Om Diagnostics}
\quad
Now we are going to consider a cosmological parameter named $Om$ \cite{Sahni3,Tong}.  It is a combination of the Hubble parameter and the cosmological
redshift,   and it provides a null test of dark energy. As a complementary to $\{r, s\}$, this diagnostic helps to distinguish the present matter density in different models more effectively.  $Om$ diagnostic has been discussed together with statefinder for many cosmological models of dark energy. It was introduced to differentiate $\Lambda$CDM from other dark energy
models. For $\Lambda$CDM model, $Om = \Omega_{m0}$ is a constant, independent
of redshift $z$. The main utility for $Om$ diagnostic is that the quantity
of $Om$ can distinguish dark energy models with less dependence
on matter density $\Omega_{m0}$ relative to the equation of state of dark energy.
For this end, we express the scale factor $a$  in terms of redshift $z$ by the relation
$1+z=a_0/a(\tau)$, where  $a_0$ is the present value of scale factor.

\begin{figure}[thbp]
\centering
\includegraphics[width=0.45\textwidth,height=0.45\textwidth]{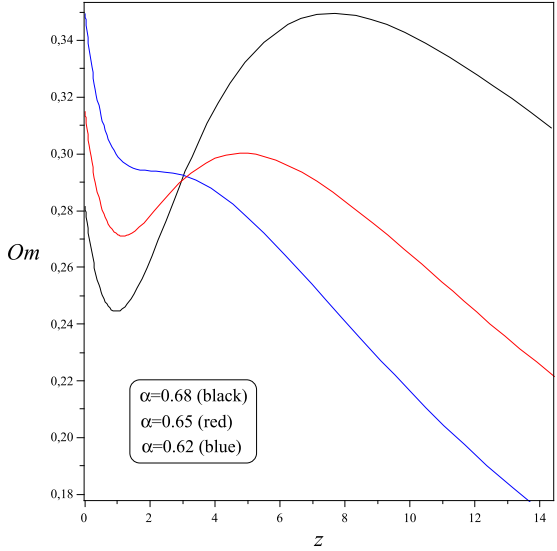}
\caption{The $Om$ parameter is plotted against the redshift $z$ for different $\alpha$, and $m=1$.}
\label{Figure_5}
\end{figure}
The starting point is the Hubble parameter which is used to determine the $Om$ diagnostic as follows:
\begin{equation}\label{22}
Om(x) = \frac{h^2(x)-1}{x^3-1},
\end{equation}
where $x=1+z, \,h(x)=H(x)/H_0$. From Eqs. (\ref{13}), (\ref{15}) and the definition (\ref{22}), we can conclude that the explicit dependence $h(x)$ can not be established. Nevertheless,  we can plot the  $Om$ curves  on $\{Om,z\}$ plane using the parametric representation
\begin{equation}
x(\tau)=\exp \left[\frac{1}{m \tau^{1-\alpha}}-\Big(1-\frac{1-\alpha}{m}\Big)\tau-\frac{2-\alpha-m}{m}\right],\nonumber
\end{equation}
\begin{equation}
h(\tau)=\left[\frac{1-\alpha}{m \tau^{2-\alpha}}+\frac{m-(1-\alpha)}{m}\right],\label{23}
\end{equation}
which is followed from (\ref{13}) and(\ref{15}).

We can find out from Fig. 5 that, at least for $\alpha >0.62$, the $Om(z)$  is of negative slope almost everywhere,
except for a certain interval of $z$. With $\alpha$ decreasing, the $Om(z)$ becomes of negative
slope for all $z$.  According to \cite{Sahni3}, it should suggest quintessence like behavior of our model ($w > -1$).

\subsection{Estimation of $\alpha$ from the $H(z)$ Data}
\quad
One could readily observe that our model contains three independent parameters namely $\alpha$, $m$ and $H_0$.  We can obtain the observational constraints on all these parameters using 28 data points of $H(z)$  in the redshift range $0.07 \le z \le 2.3$ \cite{Farooq, Busca}.
The observational data consist of measurements
of the Hubble parameter $H_{obs}(z_i)$  at redshifts $z_i$ , with
the corresponding one standard deviation uncertainties $\sigma_{Hi}$.
Following \cite{Farooq}, we define $\chi_H^2$ as
\begin{equation} \label{24}
\chi_H^2 =
\sum_{i=1}^{28}\left(\frac{H_{obs}(z_i)-H(z_i, \alpha, m, H_0)}{\sigma_{Hi}}\right)^2
\end{equation}
The observable values of $H_{obs}(z_i)$ and $\sigma_{H\,i}$ are presented in Table 1 of Ref. \cite{Farooq}, with a highest redshift measurement
at $z = 2.3$.
The problem is that we are not able to express $H(z_i, \alpha, m, H_0)$ analytically  in its explicit form as a function of $z_i$ from (\ref{13}).
At the same time, the depiction of the theoretical curve $H(z, \alpha, m, H_0)$ is not difficult, since it is possible in a parametric form $\{H(\tau), z(\tau)\} $ in accordance with (\ref{13}) and (\ref{15}). The result is presented in Fig. 6, where we take $m=1$. Moreover, one can conclude that our model with $\alpha$ around $0.68\div 0.70$ fits the observational data quite satisfactorily.

But now on, we intend to  obtain  more precise values of all parameters $\alpha, m$, and $H_0$ from  the $H(z)$ data set mentioned above. For this purpose, we consider the cosmographic fits with the well known approximation for the Hubble parameter in the form of power series in $z$
\begin{equation} \label{25}
H(z)=H_0+H_0 (1+q_0) \cdot z+\frac{1}{2} H_0 (r_0-q_0^2) \cdot z^2
+\mathcal{O}(z^3),
\end{equation}
whereas for fitting our model we employ the present values of parameters $q_0$ and $r_0$ obtained from (\ref{18}), (\ref{20}) as follows
\begin{eqnarray}
q_0 &=& q(\tau)_{\big |\tau=1}=-1+\frac{(1-\alpha)(2-\alpha)}{m},\nonumber\\
r_0 &=& r(\tau)_{\big |\tau=1}=1-\frac{\alpha(1-\alpha)(2-\alpha)}{m}.\label{26}
\end{eqnarray}

Thus, the best fit values of the model parameters from observational
Hubble data are determined by minimizing $\chi_H^2$ with $H(z; H_0, q_0, r_0)$ represented by (\ref{25}). As a result, we obtained the best fit values of parameters
as  $H_0 = 66.723179$, $q_0 = -0.4450$, and $r_0=0.6203$ with $\chi_{H\,min}^2 = 17.0719$ and the reduced $\chi_H^2$ value is $\chi^2_{red}=\chi_{H\,min}^2/\mbox{(d.o.f.)}=0.6097$ where d.o.f. is the number of degrees of freedom of that $\chi_H^2$ distribution. This shows that this model
is clearly consistent with the data since $\chi^2_{red}< 1$.

Substituting these values of $q_0$ and $r_0$  into equations (\ref{26}), we can obtain the best fit values of three parameters of our model, namely
\begin{equation} \label{27}
\alpha = 0.684,\,\,m=0.749, \,\,H_0 = 66.723 \,\, km s^{-1}Mpc^{-1},
\end{equation}
which is well consistent with the observation result from Planck+WP \cite{Planck}:
$H_0 = 67.3 ± 1.2 \,\, km s^{-1}Mpc^{-1}$.

Since the present time in our model corresponds to $\tau=1$,  the age of the Universe is approximately equal to the Hubble time, that is $t_0 = 1/H_0=14.665\,Byr$ with the same accuracy. Furthermore, using the obtained values (\ref{27}) and substituting  $q=0$ into equation (\ref{18}), we can approximately estimate  the time of ending of the first phase of accelerated expansion $\tau_e=0.030$ ($t_e=\tau_e/H_0 \approx 0.439\, Byr$) as well as the time of beginning of the present accelerated expansion $\tau_b=0.532$ ($t_b=\tau_b/H_0 \approx 7.803 \,Byr$). The observational data for $H(z)$ from Ref. \cite{Farooq}, and the model predictions (\ref{13}), (\ref{15}) are shown in Fig. 6.

\begin{figure}[thbp]
\centering
\includegraphics[width=0.45\textwidth,height=0.45\textwidth]{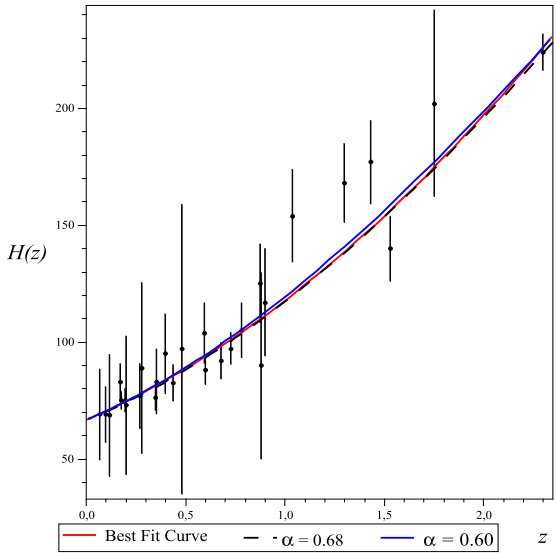}
\caption{Observational data for the Hubble parameter $H(z)$ with $1\sigma$ error bars and model
predictions versus redshift.}
\label{Figure_6}
\end{figure}

\subsection{Supernovae Type Ia Constrains}
\quad
 As is known, the SN Ia Union2 database includes 557 SNIa \cite{Amanullah, Liao} and provides one of the most powerful method for the observational restrictions on the cosmological models.
Following the maximum-likelihood approach, one should minimize $\chi^2$ and hence measure the deviations of the theoretical predictions from the observations.
Since SN Ia behave as excellent standard candles, they can be used to directly measure the expansion rate of the Universe upto high redshift, compared to the present rate. The SN Ia data allows us to measure the distance modulus $\mu$ to each supernova.
In a spatially flat Universe, the theoretical distance modulus is given by
\begin{equation}\label{28}
\mu_{th}(p_s;z) = 5 \log_{\,10} (d_L/Mpc) +25,
\end{equation}
where $d_L$ is the luminosity distance, and $p_s = \{H_0,\,m,\,n\}$ is the set of free parameters of the model to be defined through the best fit analysis.  For theoretical calculations, the luminosity distance $d_L$ of SNe Ia is defined as follows
\begin{equation}\label{29}
d_L = \frac{(1+z)}{H_0 }\int\limits_o^z \frac{d z'}{E(z')},
\end{equation}
where $E(z)=H(z)/H_0$ or $E(\tau)\equiv h(\tau)$ as it is defined by (\ref{23}).
One could choose the marginalized nuisance parameter \cite{Nesseris} for $\chi^2$
$$
\chi^2_{SNe} = A -\frac{B^2}{C}
$$
where
$$
A =\sum_{i}\Big[ \frac{\mu_{obs}(z_i)-\mu_{th}(p_s;z_i)}{\sigma_i}\Big]^2,
$$
$$
B =\sum_{i}\frac{[\mu_{obs}(z_i)-\mu_{th}(p_s;z_i)]}{\sigma_i^2},
$$
$$
C=\sum_{i}\frac{1}{\sigma_i^2},
$$
and $\sigma_i$ is the 1$\sigma$ uncertainty of the observed data $\mu_{obs}(z_i)$.

Unfortunately, we can not obtain the best-fit values of $\alpha$, $m$ and $H_0$ following the usual procedure (see, e.g., \cite{Liao} and bibliography therein). The reason is the same one we met in the previous section: one can not explicitly solve equation (\ref{15}) for $\tau_i$  from the given set of $z_i$.
With the aim of an approximate estimation of these parameters, we could take into account
the series expansion for the luminosity distance \cite{Chiba,Visser}. On purely geometrical grounds, one can obtain the luminosity distance versus redshift  as follows
\begin{equation}\label{30}
d_L \!=\! \frac{z}{H_0}\Big[1 +
\frac{1}{2}
(1 - q_0)z -\frac{1}{6}(1-q_0-3q_0^2+r_0)z^2 +O(z^3)\Big].
\end{equation}
\begin{figure}[thbp]
\centering
\includegraphics[width=0.45\textwidth,height=0.45\textwidth]{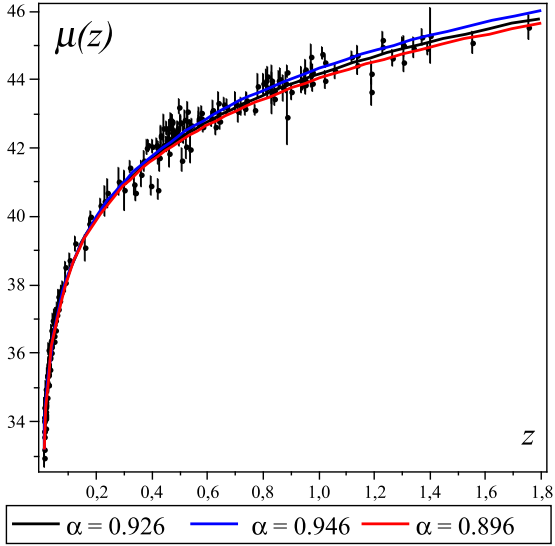}
\caption{Distance modulus in terms of the redshift for the $\Lambda_{eff} = \beta H^2$ model (m=0.174), as compared with the observational data.}
\label{Figure_7}
\end{figure}
Then, let us take the best fit values of the main cosmographic parameters in (\ref{30}) for the Union 2.1 SNIa data from Table II in Ref. \cite{Capozziello2},
\begin{eqnarray}
H_0=69.97^{+0.42}_{-0.41}\,\, km s^{-1}Mpc^{-1},\nonumber \\
q_0=-0.5422^{+0.0718}_{-0.026},\,\,r_0=0.5762^{+0.4478}_{-0.3528}.\label{31}
\end{eqnarray}
Since the age of the Universe here is approximately equal to the Hubble time, one can obtain $t_0 = 1/H_0 \approx 13.984\,Byr$. Comparing $q_0$ and $r_0 $ from Eq. (\ref{26}) with their best fit values (\ref{31}), we can obtain
\begin{equation}\label{32}
\alpha \approx 0.926,\,\,m \approx 0.174
\end{equation}
with the same accuracy as the cosmographic parameters are given in (\ref{31}).

In Fig. 7, we show a comparison between theoretical
distance modulus and observed distance modulus of supernovae
data.

\section{Conclusion}
\quad
Thus, we have tested one of the possible models  of  the fractional action cosmology which was created by means of the effective cosmological term of the particular form,  $\Lambda_{eff} = \beta H^2$. In our opinion, this model is able to describe an extremely wide period of cosmic evolution. Although our model is a phenomenological one due to its origination, it is the analytically accurate model, and therefore is quite suitable for probing the main properties of FAC. For this purpose, we have used the cosmography parameters as well as the observational data.

Nevertheless, we should note that the main purpose, we pursued in our work, consisted in the numerical evaluation of the fractional parameter, which has not been estimated in FAC earlier at all. As a result, we obtained two numerical values for $\alpha, m$ and $H_0$ given by equations (\ref{27}) and (\ref{32}). The difference in the estimates of parameters is not a surprising result. One would analyze the combined observational data, but such a result would not be so substantial. In any case, it is more important to take into account that the further study of this model outside the obtained interval for the parameters becomes leastwise speculative and leads away from reality. It is important to keep in mind that reconstructing the model by means of a scalar field or other sources, we should not get beyond the intervals obtained for the model parameters.

Of course, our result of numerical evaluation of fractional order is not a common  one, because it is highly dependent on the choice of the evolution law for the effective cosmological term. Up to date, we have not found possibility to perform any appropriate {\it universal} evaluation of the fractional order. But we do not exclude such a possibility, and we hope to consider more universal evaluation in our subsequent works.

However, once again we would like to emphasize that while developing the phenomenological cosmological models, we should  constrain the framework of the possible values of their parameters. At least in this case, one could  avoid the unwanted study of the model outside the reality.

\end{document}